\documentclass{aa}  
\usepackage{amsmath}
\usepackage{graphicx}
\usepackage{txfonts}
\usepackage{natbib}

\bibpunct{(}{)}{;}{a}{}{,} 
\usepackage{hyperref}
\hypersetup{
        colorlinks = true,
        linkcolor = blue,
        anchorcolor = red,
        citecolor = blue,
        filecolor = red,
        pagecolor = red,
        urlcolor = red
}

\def\ie{{\it i.e.,}\,}

\def\la{\hbox{\raise.5ex\hbox{$<$} 
    \kern-1.1em\lower.5ex\hbox{$\sim$}}}
\def\ga{\hbox{\raise.5ex\hbox{$>$} 
    \kern-1.1em\lower.5ex\hbox{$\sim$}}}

\newcommand{\dgr}{\mbox{$^\circ$}}           
\newcommand{\Msun}{\mbox{$M_\odot$}}         
\newcommand{\Lsun}{\mbox{$L_\odot$}}         
\newcommand{\cm}{\mbox{\ cm}}                
\newcommand{\cms}{\mbox{\ cm s${}^{-1}$}}    
\newcommand{\mes}{\mbox{\ m s${}^{-1}$}}    

\newcommand{\chem}[1]{{${}^{#1}$}} 

\begin{document}
\bibliographystyle{aa}
   \title{Multidimensional hydrodynamic simulations of the hydrogen injection flash}


   \author{M. Moc\'ak \inst{1},
           L. Siess \inst{1}
           \and E. M\"uller\inst{2}}



   \institute{Institut d'Astronomie et d'Astrophysique, Universit\'e Libre de
              Bruxelles, ULB - CP 226, 1050 Brussels, Belgium\\
              \email{mmocak@ulb.ac.be, siess@astro.ulb.ac.be}         
          \and Max-Planck-Institut f\"ur Astrophysik,
              Postfach 1312, 85741 Garching, Germany\\
              \email{ewald@mpa-garching.mpg.de}}

   \date{Received  ........................... }


\abstract
{The injection of hydrogen into the convection shell powered by helium
  burning during the core helium flash is commonly encountered during
  the evolution of metal-free and extremely metal-poor low-mass stars.
  Recent multidimensional hydrodynamic simulations indicate that
  the hydrogen injection may also occur in more metal-rich stars due to
  turbulent entrainment which accelerates the growth of the shell
  convection zone and increases its size. However, 1D stellar models
  cast doubts that helium-flash hydrogen mixing does occur as it
  requires the crossing of an entropy barrier at the helium-hydrogen
  interface.  }
%
{With specifically designed multidimensional hydrodynamic simulations,
  we aim to prove that an entropy barrier is no obstacle for the
  growth of the helium-burning shell convection zone in the helium
  core of a metal-rich Pop I star, \ie convection can penetrate into
  the hydrogen-rich layers for these stars, too. We further study
  whether this is also possible in one-dimensional stellar
  evolutionary calculations}
%
{We artificially shift the hydrogen-rich layer closer to the outer
  edge of the helium-burning shell convection zone in a Pop I star
  with a mass of 1.25~\Msun, and simulate the subsequent evolution in
  2D and 3D, respectively. We also perform stellar evolutionary
  calculations of the core helium flash in metal-rich stars
  implementing turbulent entrainment by means of a simple
  prescription. These simulations were performed with the Eulerian
  hydrodynamics code HERAKLES and the stellar evolution code STAREVOL,
  respectively.}
%
{Our hydrodynamical simulations show that the helium-burning shell
  convection zone in the helium core moves across the entropy barrier and
  reaches the hydrogen-rich layers. This leads to mixing of protons into
  the hotter layers of the core and to a rapid increase of the nuclear 
  energy production at the
  upper edge of the helium-burning convection shell - the hydrogen
  injection flash. As a result a second convection zone appears in the
  hydrogen-rich layers. Contrary to 1D models, the entropy barrier
  separating the two convective shells from each other is largely permeable
  to chemical transport when allowing for multidimensional flow, and
  consequently, hydrogen is continuously mixed deep into the helium
  core. We find it difficult to achieve such a behavior in one-dimensional
  stellar evolutionary calculations.}
{}


\maketitle


\section{Introduction}
\label{sect:intro}

The hydrogen injection flash is believed to commence during the core helium
flash of metal-free and extremely metal-poor (metallicity Z $<10^{-4}$)
low-mass stars when their helium core shell convection zone penetrates into
the hydrogen-rich layers \citep{FujimotoIben1990, Hollowell1990,
SchlattlCassisiSalaris2001, CassisiSchlattl2003, WeissSchlattl2004,
Campbell2008}. Such an injection of protons is not expected for more
metal-rich low-mass stars because the ignition of helium occurs further
away from the hydrogen-rich layer and the entropy barrier between the
helium and hydrogen layers is higher.

\begin{figure*} 
\includegraphics[width=0.49\hsize]{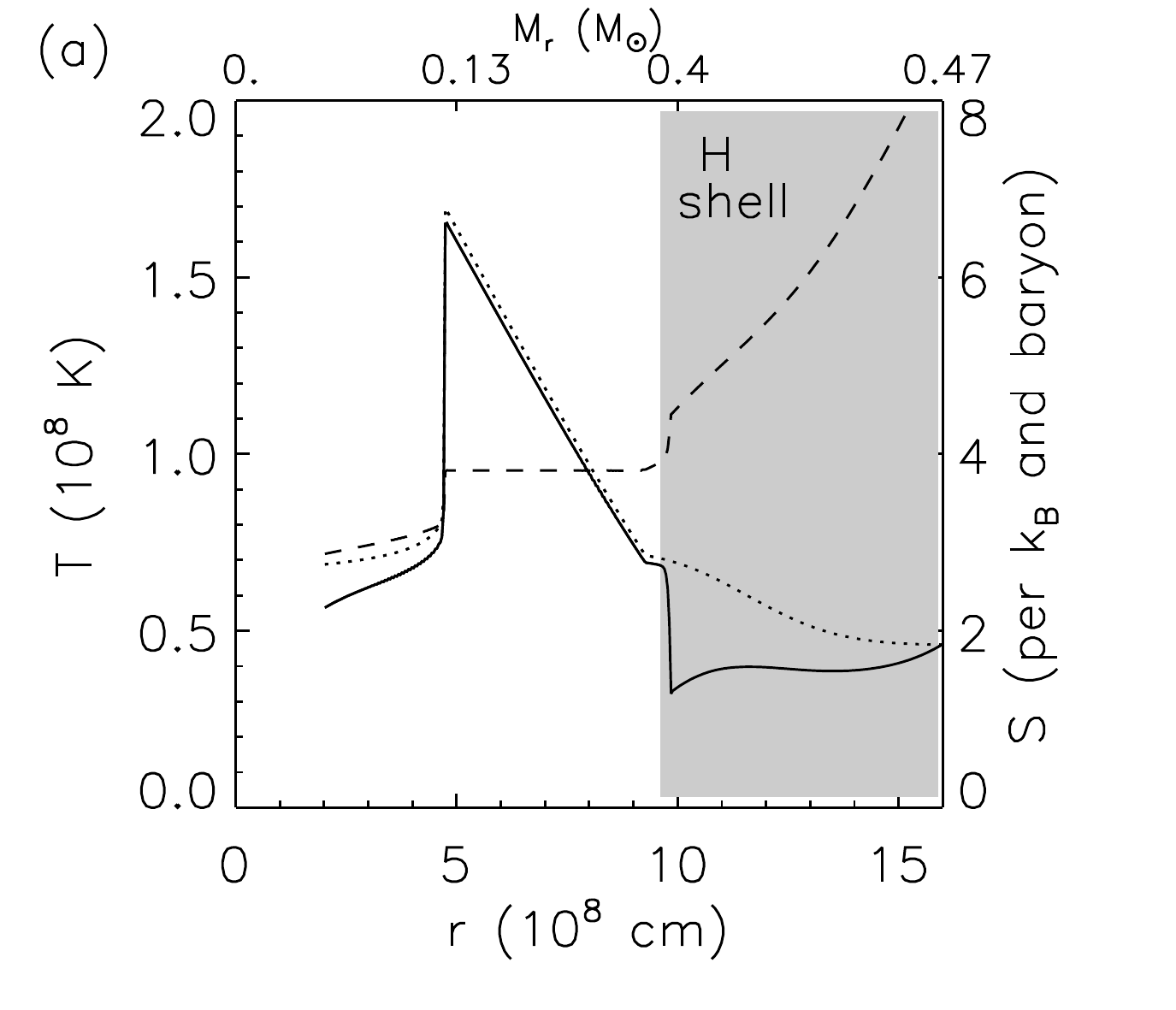}
\includegraphics[width=0.49\hsize]{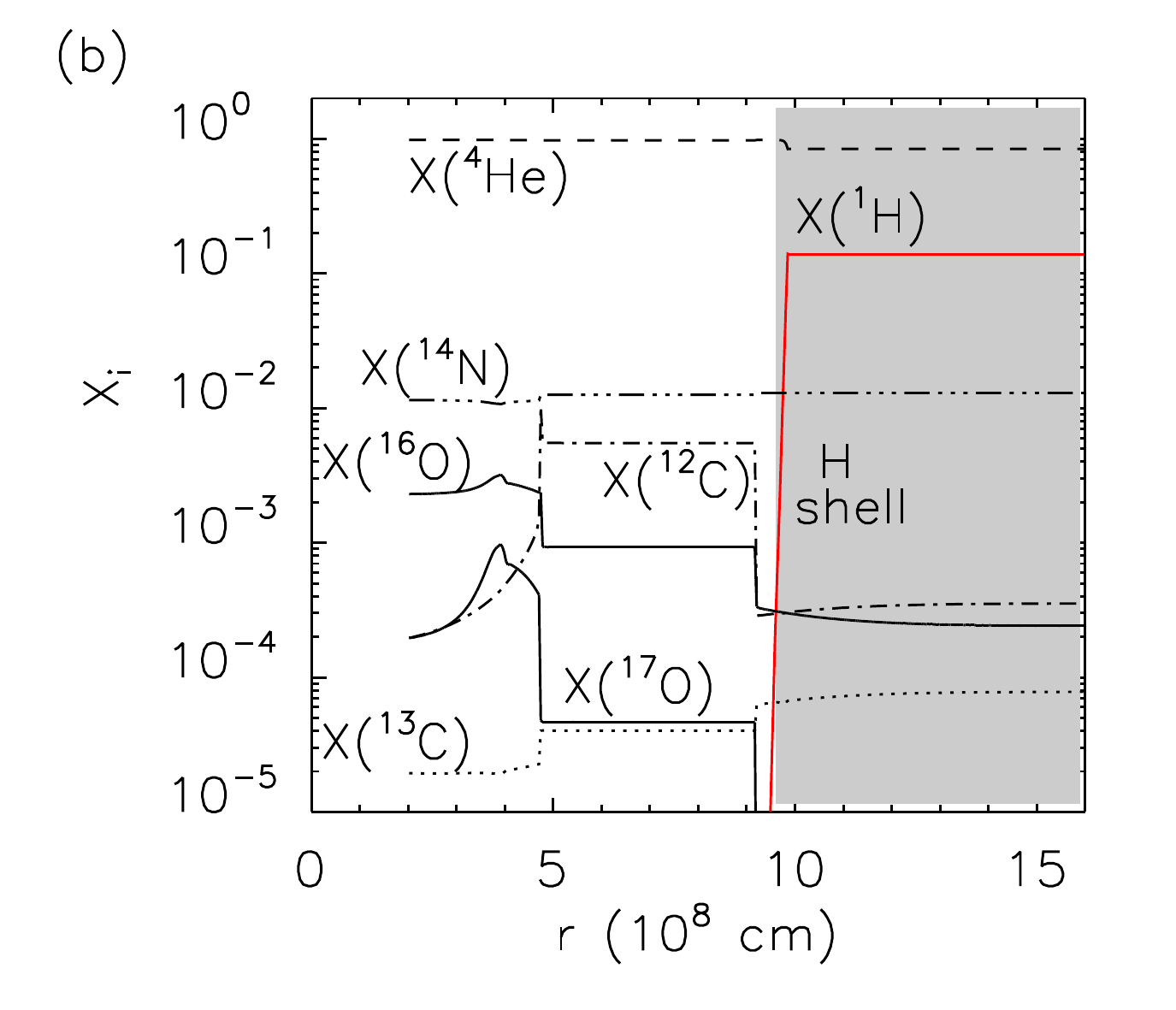}
\caption{(a; left panel): Temperature distribution of the modified
  initial model (MOD, solid) and of the original model (M, dotted) as
  a function of radius $r$ and of enclosed mass M$_r$ in solar units
  M$_\odot$, respectively. The dashed curve gives the entropy profile
  of model MOD. (b; right panel) Mass fraction of the main isotopes of
  model MOD as a function of radius. The hydrogen shell corresponds to
  the shaded region.}
\label{fig.inimod}
\end{figure*}

Multidimensional hydrodynamic simulations demonstrate that entropy
barriers in stars are not as impenetrable as they may seem to be at
first sight. The mixing beyond conventional Schwarzschild or Ledoux
convection boundaries and the growth of convection zones on
dynamic timescales due to turbulent entrainment are proves of it
\citep{MeakinArnett2007}. Recent hydrodynamic simulations of shell
convection during the core helium flash also suggest that convection
could reach the outer hydrogen-rich layers of the star within a week
\footnote{We find that the growth rate of the shell convection due to
  turbulent entrainment is of the order of a few meters per second for
  our hydrodynamic models, the exact value depending on the stability
  of the convection zone boundaries and the convective velocities near
  these boundaries \citep{MeakinArnett2007}. The latter in turn depend
  not only on dimensionality of the flow (2D, 3D), but also on the
  spatial resolution of the simulation.}
at around the peak of the core helium flash even in metal-rich stars
\citep{Mocak2009,Mocak2010}. However in these models, the
hydrogen-rich layers are still far away from the helium-burning shell
convection zone preventing the simulation of the expected proton
injection process within an affordable amount of computing time.
Alternatively, one may study the evolution during the core helium
flash up to a possible hydrogen injection by one-dimensional stellar
evolutionary calculations which include the effects of turbulent
entrainment by means of a simple entrainment law. As such simulations
are still missing, we implemented in our 1D stellar evolution code a
simple entrainment law based on the ideas of \citet{MeakinArnett2007},
and performed stellar evolutionary calculations of the core helium
flash in low-mass Pop I stars. The entrainment rates adopted for these
calculations are taken from three-dimensional hydrodynamic simulations
of \citet{Mocak2010}.

Hydrodynamic simulations of similar phases of stellar evolution are
currently also carried out by \citet{Herwig2006,Herwig2011}. The
results of their and our studies can have important implications for
the hydrogen injection not only during the core helium flash but also
during the early-AGB phase (\ie the ``dual shell flashes'' described
in \citet{Campbell2008}) or in more advanced nuclear burning stages.

The paper is organized as follows. We introduce the numerical tools
and the stellar input model used for our simulations in the next
section. The hydrodynamic simulations of the hydrogen injection phase
and the stellar evolutionary calculations mimicking turbulent
entrainment during the core helium flash are presented in
Sect.\,\ref{sect:hydrosim}, and Sect.\,\ref{sect:stec},
respectively. A summary of our findings is given in Sect.\,\ref{sect:sum}.

\begin{figure} 
\includegraphics[width=0.89\hsize]{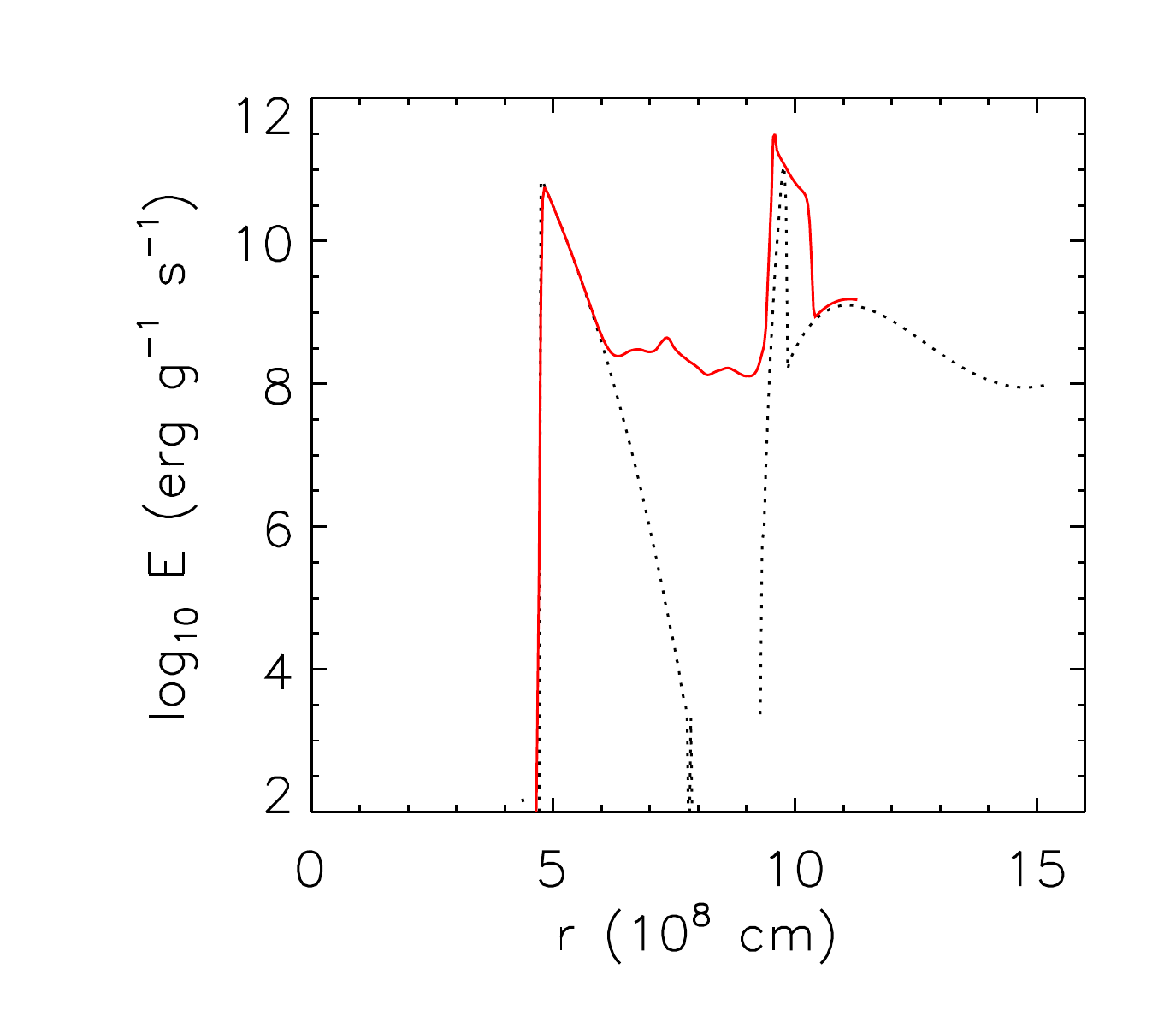}
\caption{Nuclear energy production rate as a function of radius $r$
  for the 3D model exphif.3d.f at $t = 0\,$s (dotted-black),
  and at $t = 9800\,$s (solid-red), respectively.}
\label{fig.egen}
\end{figure}

\section{Physical Input and Codes}
\label{sect:inpcod}

\begin{table} 
\caption{Some properties of the modified initial stellar model MOD:
  Total mass $M$, mass $M_{He}$ and radius $R_{He}$ of the helium
  core, total nuclear energy production by helium burning $L_{He}$,
  and by hydrogen burning $L_H^p$ and $L_H^f$, where the superscripts
  $p$ and $f$ denote that the respective numbers were obtained using a
  partial and full CNO network, respectively.}
\center{
\begin{tabular}{p{0.8cm}|p{0.5cm}p{0.5cm}p{1.cm}p{1.cm}p{1.cm}p{1.cm}
p{0.8cm}} 
\hline
\hline
Model & $M$  & $M_{He}$  & $R_{He}$ & $L_{He}$ & $L_{H}^p$ & $L_{H}^f$ \\ 
& $[\Msun]$ &  $[\Msun]$ & $[10^8\cm]$ & $[10^9\Lsun]$  & 
$[10^9\Lsun]$ & $[10^9\Lsun]$ \\
\hline
MOD  & $1.25$ & $0.4$ & 9.6 & 0.6 & 0.1 & 0.3 \\
\hline
\end{tabular} 
}
\label{tab.inim} 
\end{table} 

For the multidimensional hydrodynamic simulations we used the HERAKLES
code \citep{Mocak2008} derived from the PROMETHEUS code
  \citep{FryxellArnettMueller1991, MuellerFryxellArnett1991}. In
  brief, HERAKLES solves the Euler equations coupled with source terms
  describing self-gravity and nuclear burning. The hydrodynamic
  equations are integrated with the PPM reconstruction scheme
  \citep{ColellaWoodward1984} and a Riemann solver for real gases
  according to \citet{ColellaGlaz1984}. The evolution of the chemical
  species is described by a set of additional continuity equations
  \citep{PlewaMueller1999}. Self-gravity is handled according to
  \citet{MuellerSteimnetz1995}, gravitational potential being
  approximated by a 1D Newtonian potential obtained from the
  spherically averaged mass distribution. The code integrates nuclear
  networks with the semi-implicit Bader-Deufelhard method
  \citep{BaderDeuflhard1983,Press1992}. 

\begin{table*} 
\caption{Some properties of the 2D and 3D hydrodynamic simulations
  based on model MOD: wedge size $w$; number of grid points and
  resolution in $r$, $\theta, and \,\phi$ direction, respectively;
  estimated Reynolds number R$_e$; characteristic velocity v$_{c}$ of
  the convective flow in helium-burning layers; typical convective
  turnover timescale $\tau_{conv}$ at time at $t \sim 40\,000\,$s for
  model 2d.p, and at $t \sim 2500\,$s for models 2d.f and 3d.f,
  respectively; $t_{max}$ final evolutionary time.}
\begin{center}
\begin{tabular}{p{1.5cm}|p{0.7cm}p{1.6cm}p{1.3cm}p{0.5cm}p{0.5cm}
p{0.5cm}p{1.4cm}p{1.cm}p{1.3cm}} 
\hline
\hline
run & $w$ & grid & $\Delta r$ & $\Delta\theta$ & $\Delta\phi$ & R$_e$ &  
$v_{c}$ & $\tau_{conv}$ & $t_{max}$ \\
$\mbox{[name]}$ & [$\dgr$] & N$_r \times$N$_\theta \times$N$_\phi$  & 
[10$^{6}$ cm] & [$\dgr$] & [$\dgr$] & & [10$^{6}\,\cms$]  & [s] & [s]\\
\hline 
hifexp.3d.f & 45 & $370\times45\times45$ & 3.8 & 1 & 1 & $10^2$ & 1.0 & 1000 & 14400 \\
hifexp.2d.f & 90 & $370\times90$ & 3.8 & 1 & - & $10^2$ & 1.7 & 600 & 14400 \\
hifexp.2d.p & 90 & $370\times90$ & 3.8 & 1 & - & $10^2$ & 1.8 & 550 & 200000 \\
\hline
\end{tabular} 
\end{center}
\label{tab.modhyd} 
\end{table*}  

As an initial model we used the helium core structure of a
1.25\,\Msun, $Z=Z_\odot$ star during the core helium flash at its peak
nuclear luminosity (Tab.\,\ref{tab.inim}). In the initial model,
computed with the GARSTEC stellar evolution code
\citep{WeissSchlattl2000,WeissSchlattl2007}, the hydrogen-rich layers
are artificially shifted closer to the outer boundary of the
helium-burning shell convection zone at radius around $9.8 \times
10^8\,$cm.  Originally the lower boundary of the hydrogen layer is
located at a radius of $1.9\times 10^9\,$cm. The unmodified initial
model (M) is described in more detail in
\citet{Mocak2008,Mocak2009}. Compared to this model the resulting
hydrostatic structure of the modified initial model MOD differs by a
temperature sink at the boundary between the helium and hydrogen-rich
layers (H-He boundary) at $9.5 \times 10^8\,\mbox{cm} < r < 10\times
10^8\,$cm (Fig.\,\ref{fig.inimod}). At this boundary the entropy
increases almost discontinuously by 20$\%$, producing the so-called
barrier. The temperature gradient in the helium-burning convection
zone ($4.7 \times 10^8\,\mbox{cm} < r < 9.2 \times 10^8\,$cm) remains
almost unaltered as well as the nuclear energy production.

\begin{figure*} 
\includegraphics[width=0.49\hsize]{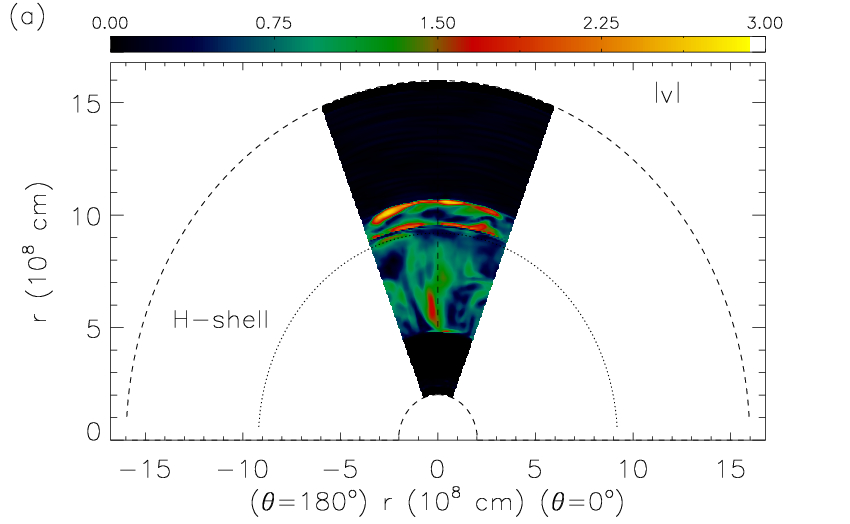}
\includegraphics[width=0.49\hsize]{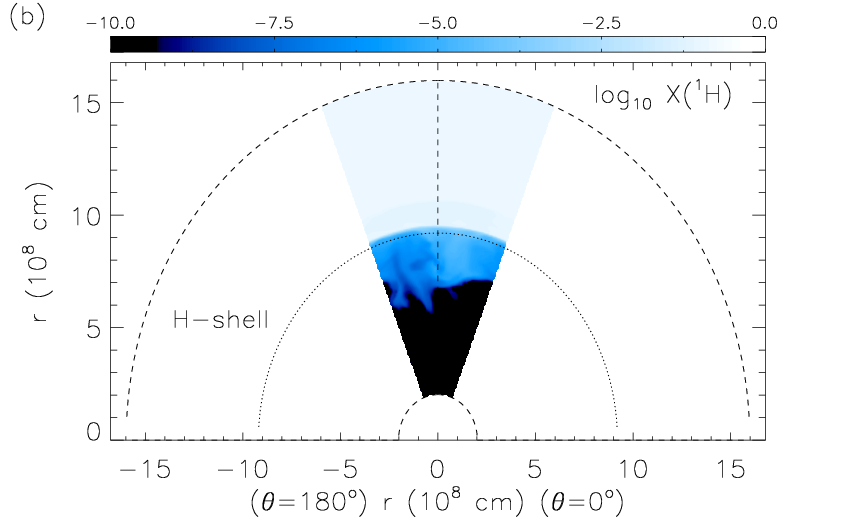}
\includegraphics[width=0.49\hsize]{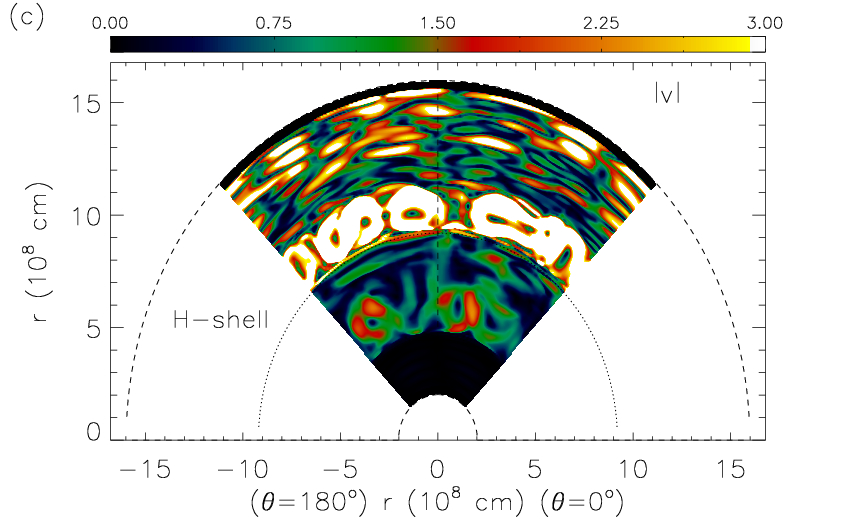}
\hspace{0.2cm}
\includegraphics[width=0.49\hsize]{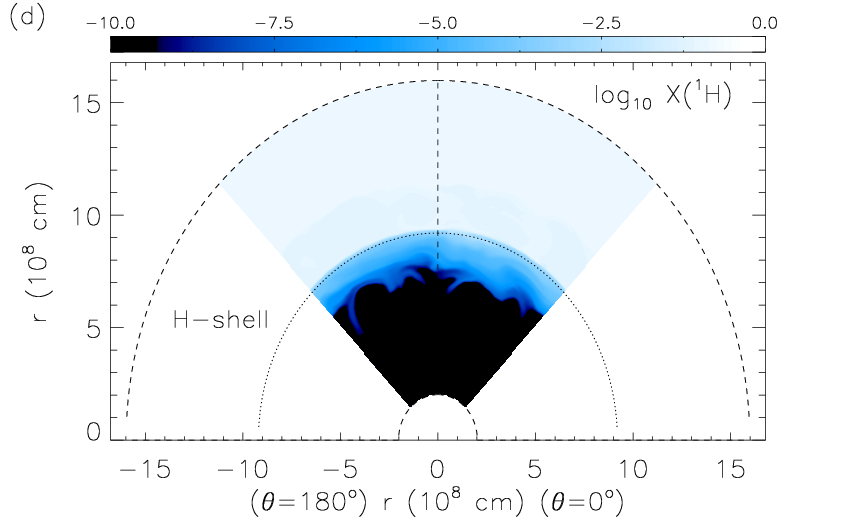}
\caption{Top row: snapshots showing contour plots of the modulus of
  the velocity $|$v$|$ (in units of $10^6 \cms$; left panel) and the
  hydrogen mass fraction (right panel) for the 3D model hifexp.3d.f at
  t$\sim 13300\,$s in the meridional plane ($\phi = 0\dgr$). Bottom
  row: same quantities but for the 2D model hifexp.2d.f. The dotted
  line separates helium- and hydrogen-rich layers (H-shell), and the dashed
  lines mark the boundaries of the computational domain.}
\label{fig.snap3d2dhif2}
\end{figure*}

At the H-He boundary of model MOD around $r \sim 9.8\times 10^8\,$cm,
we find a narrow peak in the nuclear energy production rate
(Fig.\,\ref{fig.egen}) that is due to some left-over hydrogen ($X
~\sim 10^{-4}$) when modifying the original initial model. In the
layers near the H-He boundary the temperature is relatively high ($T
\sim 5\times 10^7\,$K), and even if the hydrogen mass fraction is
small, the nuclear energy production is significant. Although, this
feature is not seen in 1D stellar evolution simulations of the
hydrogen injection flash \citep{Hollowell1990}, it may represent a
somewhat more ``realistic'' situation, where the H-He interface is not
a discontinuity but a moderately hot layer partially enriched with
hydrogen.

For the stellar evolutionary calculations of the core helium flash we
implemented a simple turbulent entrainment law in the STAREVOL code
\citep{Siess2006} and evolved a 1~\Msun, $Z = Z_\odot$ star for
different entrainment rates guided by our 3D hydrodynamic simulations.
To achieve hydrogen injection even in metal-rich low-mass stars, we
also tried the classical overshooting prescription based on the
exponential decay of the convective velocity field beyond the
convection zone boundaries \citep{Freytag1996} with various
efficiencies.

STAREVOL is a 1D lagrangian stellar evolution code able to
  compute the evolution of stars in the mass range (0.1 - 60
  $M_\odot$) from pre-main sequence up to Neon ignition. The nuclear
  network includes 53 species coupled by 177 nuclear reactions
  \citep{SiessArnould2008}, an accurate equation of state tested in
  the domain of very low mass stars \citep{SiessDufour2000}, and all
  relevant neutrino contributions (for a detailed description, the
  reader is referred to \citet{Siess2006}. This code has been widely
  used to study pre-main sequence stars \citep{SiessForestini1997},
  rotational mixing \citep{Palacios2006}, AGB nucleosynthesis
  \citep{SiessLivio2002} and super AGB stars \citep{Siess2010}.

\section{Hydrodynamic simulations}
\label{sect:hydrosim}

We performed two 2D (models hifexp.2d.f and hifexp.2d.p) and one 3D
(model hifexp.3d.f) hydrodynamic simulation using different nuclear
reaction networks. The last letter of the model name, $f$ and $p$,
refers to simulations performed with a full and partial CNO reaction
network (\ie neglecting reactions related to $^{13}$N and
$^{15}$O). These runs start from the modified initial model MOD with
the shifted hydrogen profile. Some general properties of these
simulations are summarized in the Tab.\,\ref{tab.modhyd}, all of which
were performed on an equidistant spherical grid ranging in radius from
$2\times 10^8\cm$ to $1.6\times 10^9\cm$. The adopted grid
  resolution is motivated by our earlier work \citep{Mocak2008}, where
  we found convergence of the results starting at a radial grid
  resolution $\Delta r \sim 3.7 \times 10^6\,$cm and an angular
  resolution of $\sim 1\dgr$.  The boundary conditions in the radial
direction are reflective, while periodic boundaries are imposed in the
angular directions. For models hifexp.3d.f and hifexp.2d.f the
reaction network includes the following species: $^1$H, $^3$He,
$^4$He, $^{12}$C, $^{13}$C, $^{13}$N, $^{14}$N, $^{15}$N, $^{15}$O,
$^{16}$O, $^{17}$O, $^{20}$Ne, $^{24}$Mg, and $^{28}$Si, coupled by
reactions of the full CNO cycle and the triple-$\alpha$ reaction. 
  More details are given in the PhD thesis of \citet{Mocak2009phd}.

For the simulation hifexp.2d.p we omitted the isotopes $^{13}$N and
$^{15}$O, the $\beta-$decay reactions of the CNO cycle and the
$^{12}$C(p,$\gamma$)$^{13}$N reaction. Although this variant is
unrealistic, it allows us to demonstrate that entropy barriers which
are not sustained by sufficiently strong nuclear burning can be
completely destroyed by the growing convection zone due to turbulent
entrainment. When considering all species of the full CNO cycle, the
nuclear energy production at the H-He boundary is a factor of 3 higher
than in the partial network and leads to the formation of a shell
convection zone in the hydrogen-rich layers
(Fig.\,\ref{fig.snap3d2dhif2}; models hifexp.2d.f and hifexp.3d.f; see
Sect.\,\ref{sect.simfullcno}). No shell convection zone appears in the
hydrogen-rich layers of the 2D model hifexp.2d.p.

\subsection{Simulations with the full CNO reaction network}
\label{sect.simfullcno}

Our hydrodynamic 3D and 2D models, hifexp.3d.f and hifexp.2d.f
(Fig.\,\ref{fig.snap3d2dhif2}), are evolved with an accurate treatment
of the CNO cycle on a numerical grid having the same number of zones
in radial and angular directions. The latter allows us to compare the
results of these simulations being biased only by the dimensionality
of the simulation, but not by the grid resolution.

\begin{figure} 
\includegraphics[width=0.99\hsize]{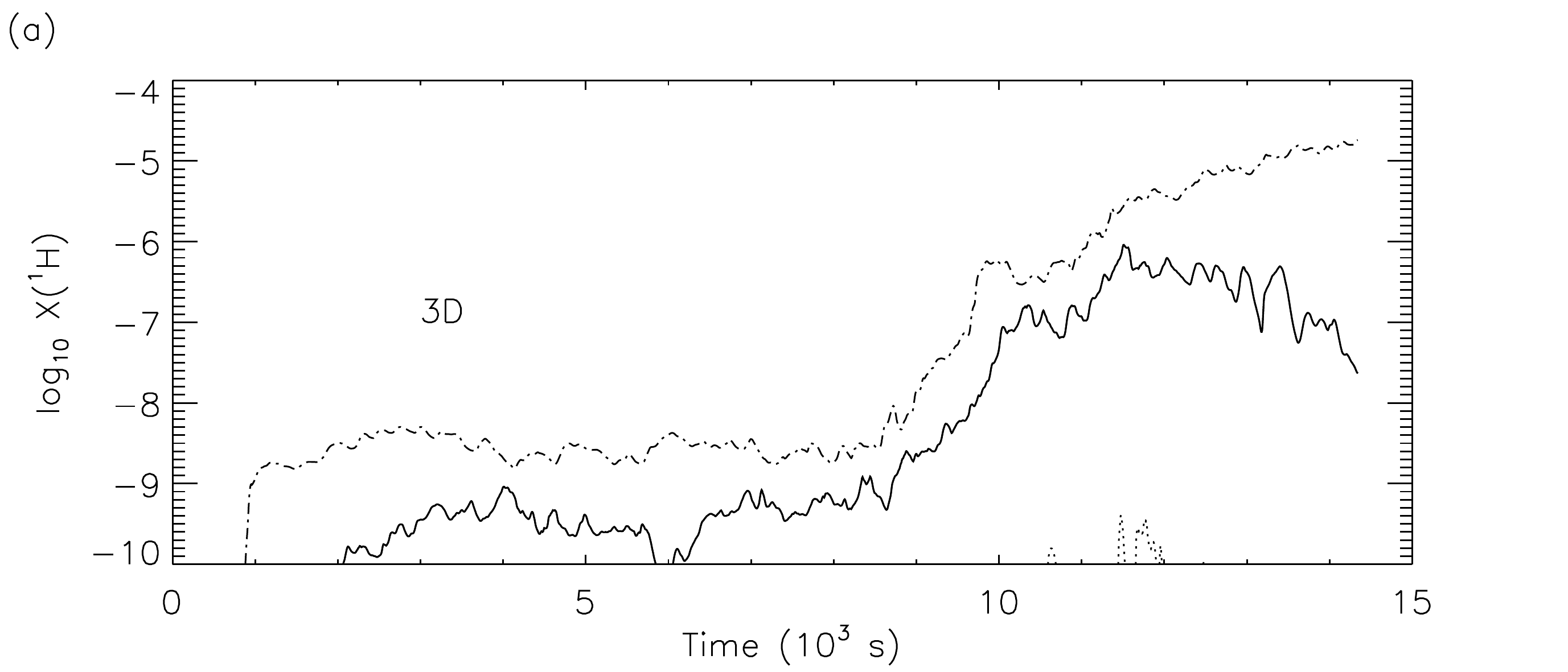}
\includegraphics[width=0.99\hsize]{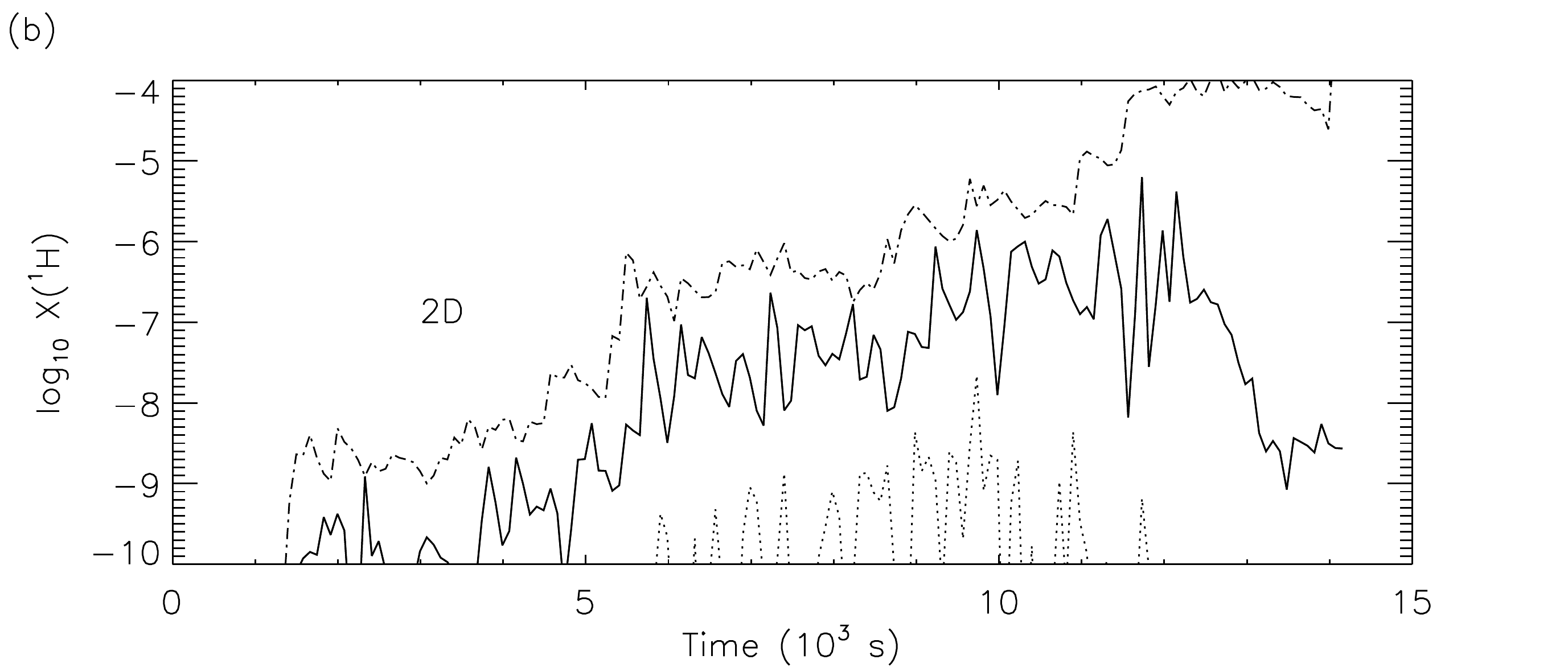}
\caption{Temporal evolution of the (logarithm of the) hydrogen mass
  fraction X($^1$H) at three different radii [temperatures] around the
  top, middle, and bottom of the helium-burning convection shell,
  respectively: $r_1 = 8.8\times 10^8\,$cm [$T_1 \sim 0.8\times
    10^8\,$K] (dash-dotted), $r_2 = 7.1\times 10^8\,$cm [$T_2 \sim
    1.2\times 10^8\,$K] (solid), and $r_3 = 5.4\times 10^8\,$cm [$T_3
    \sim 1.5 \times 10^8\,$K] (dotted). The upper panel (a) gives the
  results for the 3D model hifexp.3d.f, and the lower panel (b) for
  the 2D model hifexp2d.f, respectively.}
\label{2d3ddredgedown}
\end{figure}

In the 3D (2D) model, hifexp.3d.f (hifexp.2d.f), after 800\,s
(1000\,s) convection fully develops in the layers located above the
temperature maximum $T_{max}$ ($r \sim 4.7 \times 10^8\,$cm). As soon
as the first convective plumes reach the H-He boundary, the
dredge-down of hydrogen sets in.  Up to $\sim 1500\,$s the composition
profile in both models is similar, with a steep increase of the
hydrogen mass fraction X($^1$H) exceeding 10$^{-9}$ just below the
H-He interface (Fig.\,\ref{2d3ddredgedown}). Subsequently and until
$t\sim 8000\,$s, X($^1$H) remains almost constant in the
helium-burning convection zone of the 3D model, where
$10^{-10}<$~X($^1$H)$~<~10^{-8}$. This is in contrast with the 2D
simulation, hifexp.2d.f, where X($^1$H) continues to increase during
this epoch. Between 8000\,s and 12\,000 s, the proton mass fraction
increases in the helium-burning convection of both 2D and 3D models.

\begin{figure} 
\includegraphics[width=0.99\hsize]{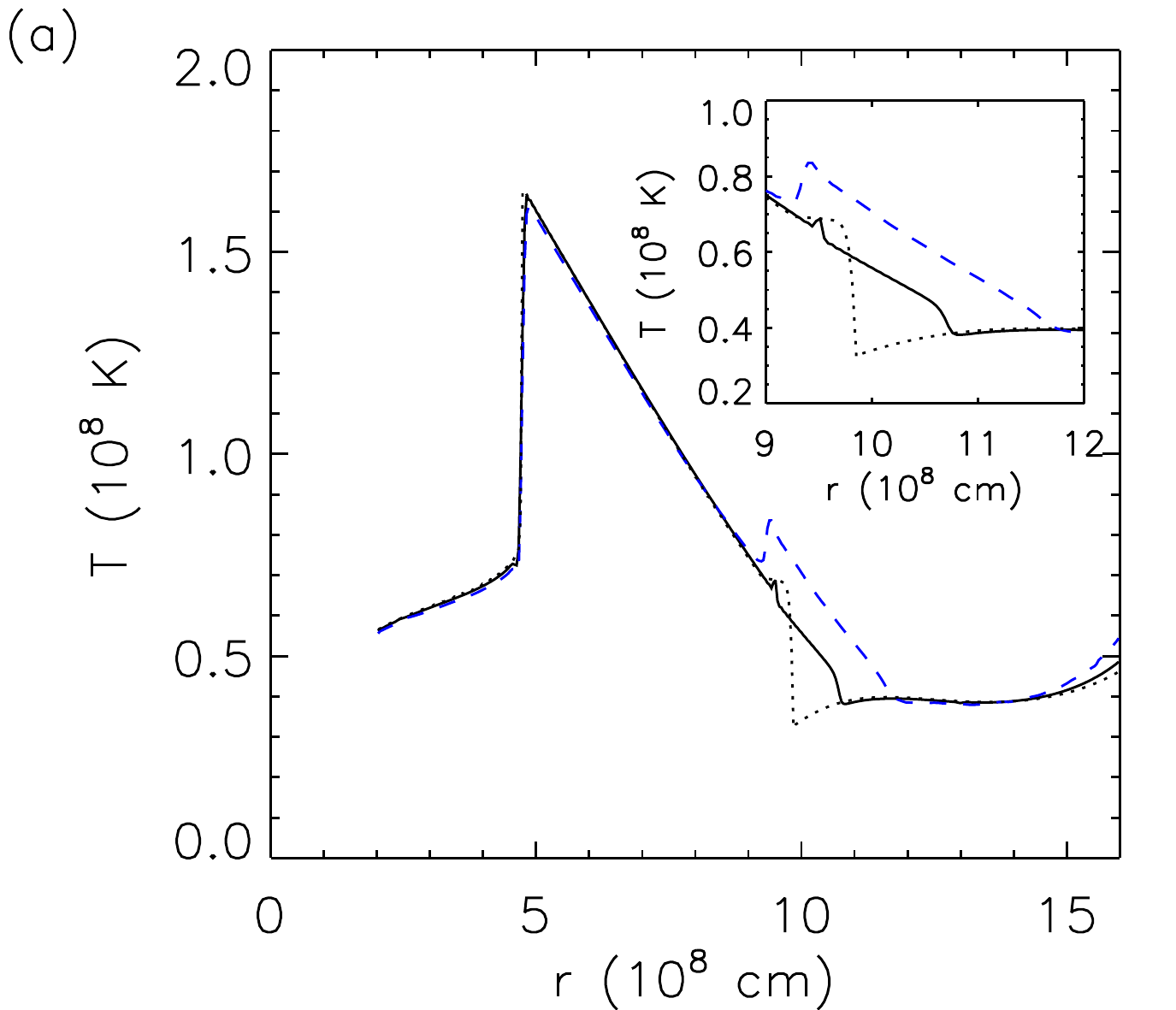}

\includegraphics[width=0.99\hsize]{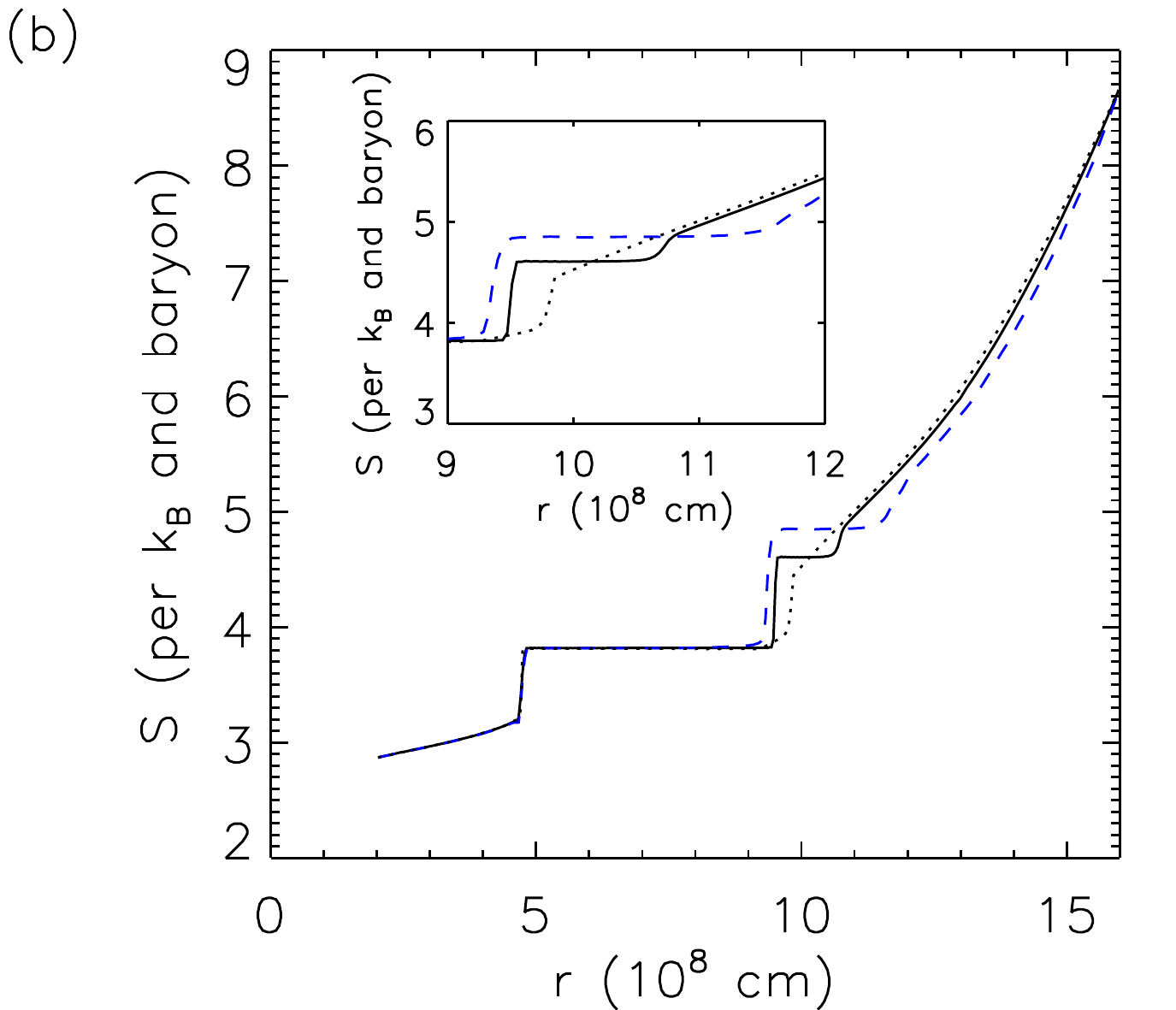}
\caption{Radial distribution of (a) temperature and (b) entropy at $t
  = 0\,$s (dotted) and $t_1 = 13\,500\,$s for the 2D model hifexp.2d.f
  (dashed, blue), and the 3D model hifexp.3d.f (solid),
  respectively. The inserts show a zoom of the region around the H-He
  interface.}
\label{fig.temph1entr}
\end{figure}

At $t\sim 12\,000\,$s, the entropy barrier at the H-He interface in
both models is steeper and almost twice as large as it was at the
beginning of the simulations due to the enhanced nuclear burning. This
increase in temperature and entropy at the interface slows down the
injection of protons and the hydrogen mass fraction in the helium-rich
layers decreases (Fig.\,\ref{2d3ddredgedown}).

As protons are dredged down into the helium core, they are captured by
$^{12}$C via the $^{12}\mbox{C}(\mbox{p},\gamma)^{13}\mbox{N}$
reaction, \ie the hydrogen abundance decreases sharply with
depth. Nevertheless, some protons can reach layers with temperatures
exceeding $10^8\,$K, where their lifetime against the
$^{12}\mbox{C}(\mbox{p}, \gamma)^{13}\mbox{N}$ reaction is $\lesssim
100\,$s \citep{Caughlan1988}. This is much shorter than their lifetime
outside the He-burning convection shell. As the convective plumes have
velocities of $10^6 \cms$ ($2\times 10^6 \cms$) in the 3D (2D) model,
hydrogen-rich gas is advected from the ``colder'' top of the
He-burning convection shell to the aforementioned hot layers (located
at $r \sim 7.7\times 10^8 \cms$) in less then 200\,s (100\,s).  Our
simulations show that this time is too short for the hydrogen to burn
completely while being advected to layers with temperatures
$\gtrsim\,10^8\,$K (Fig.\,\ref{2d3ddredgedown}).

The differences between the X($^1$H) distributions of the 2D and 3D
model can be understood by realizing that the convective velocities in
the 3D model are almost a factor of two smaller than in the 2D one.
Hence, there is more time for hydrogen to burn during its transport in
the 3D model than in the 2D one, \ie hydrogen is able to reach the hot
layers in larger amounts in the latter case
(Fig.\,\ref{2d3ddredgedown}).  Besides higher convective velocities,
the convective plumes arising at the H-He boundary also fill a larger
volume fraction in the 2D model compared with the 3D one, where the
plumes are smaller and narrower.

The nuclear energy production at the H-He boundary and the impacting
convective plumes change the initial properties of the H-He
interface. At the start of the simulations, the boundary is
characterized by a steep radial step in both the temperature and
entropy profiles. However, when convection reaches the interface, the
single step structure of the temperature and entropy profiles evolves
into a two steps structure exhibiting a narrow entropy plateau between
the steps (Fig.\,\ref{fig.temph1entr}). This plateau reflects the
formation of a secondary convection shell, located in the
hydrogen-rich layers and powered by the CNO cycle. We find higher
absolute values of both the entropy and temperature in the 2D model,
hifexp.2d.f, compared to its 3D counterpart across the forming new
convection shell. This can be explained by the stronger nuclear
burning occurring in that model, which also leads to a faster growth
of the hydrogen-burning convection shell.

\begin{figure*} 
\includegraphics[width=0.33\hsize]{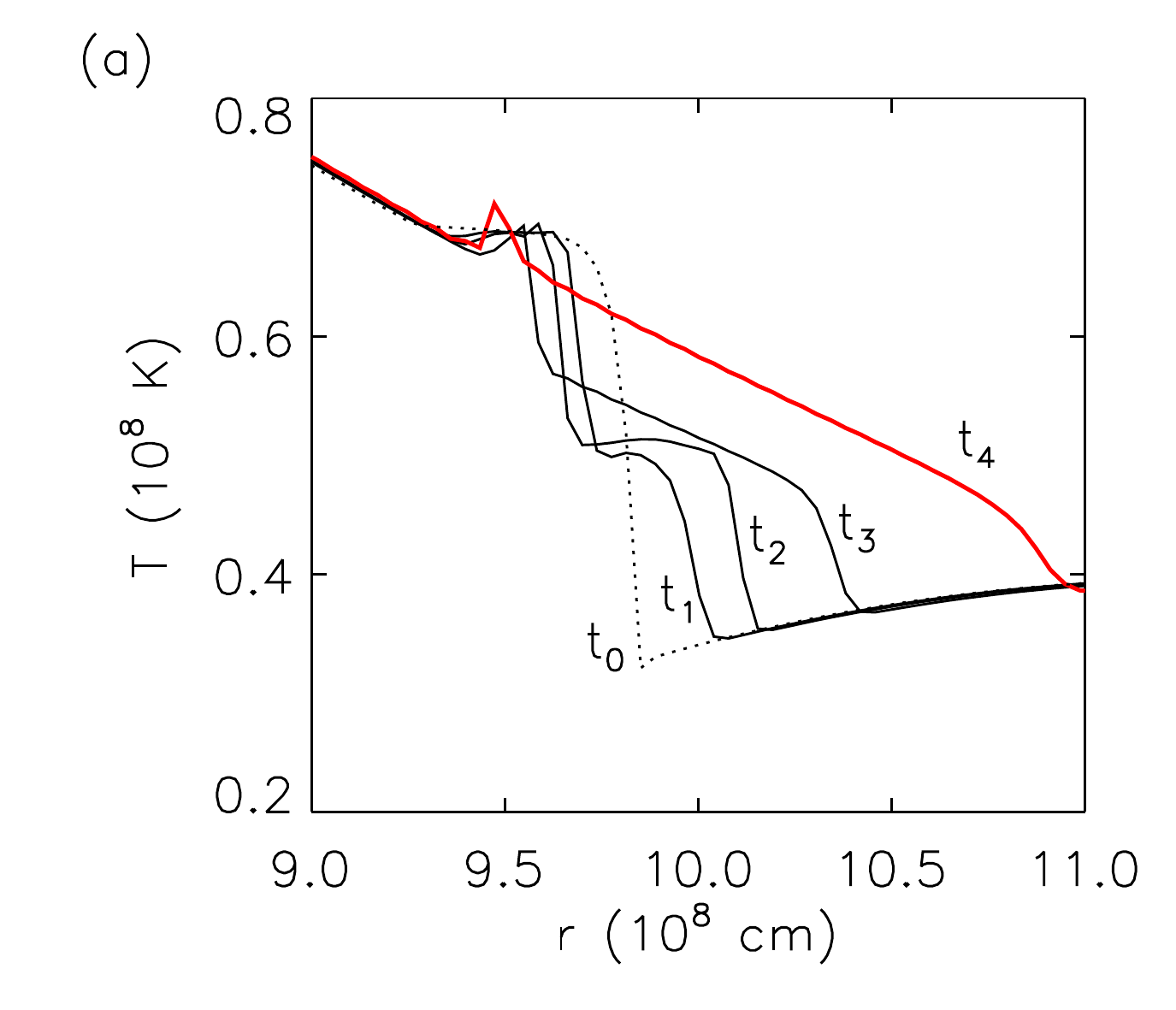}
\includegraphics[width=0.33\hsize]{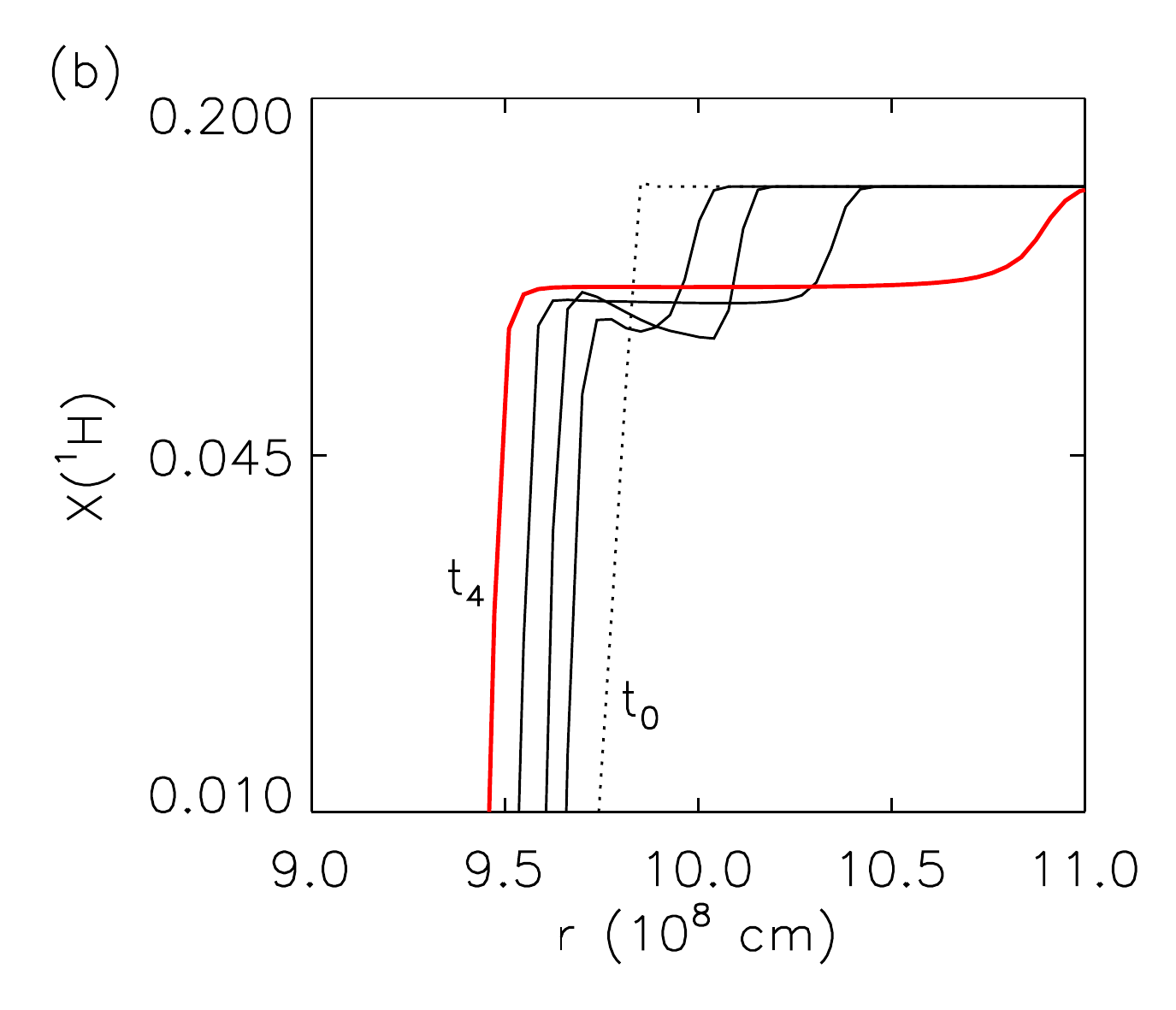}
\includegraphics[width=0.33\hsize]{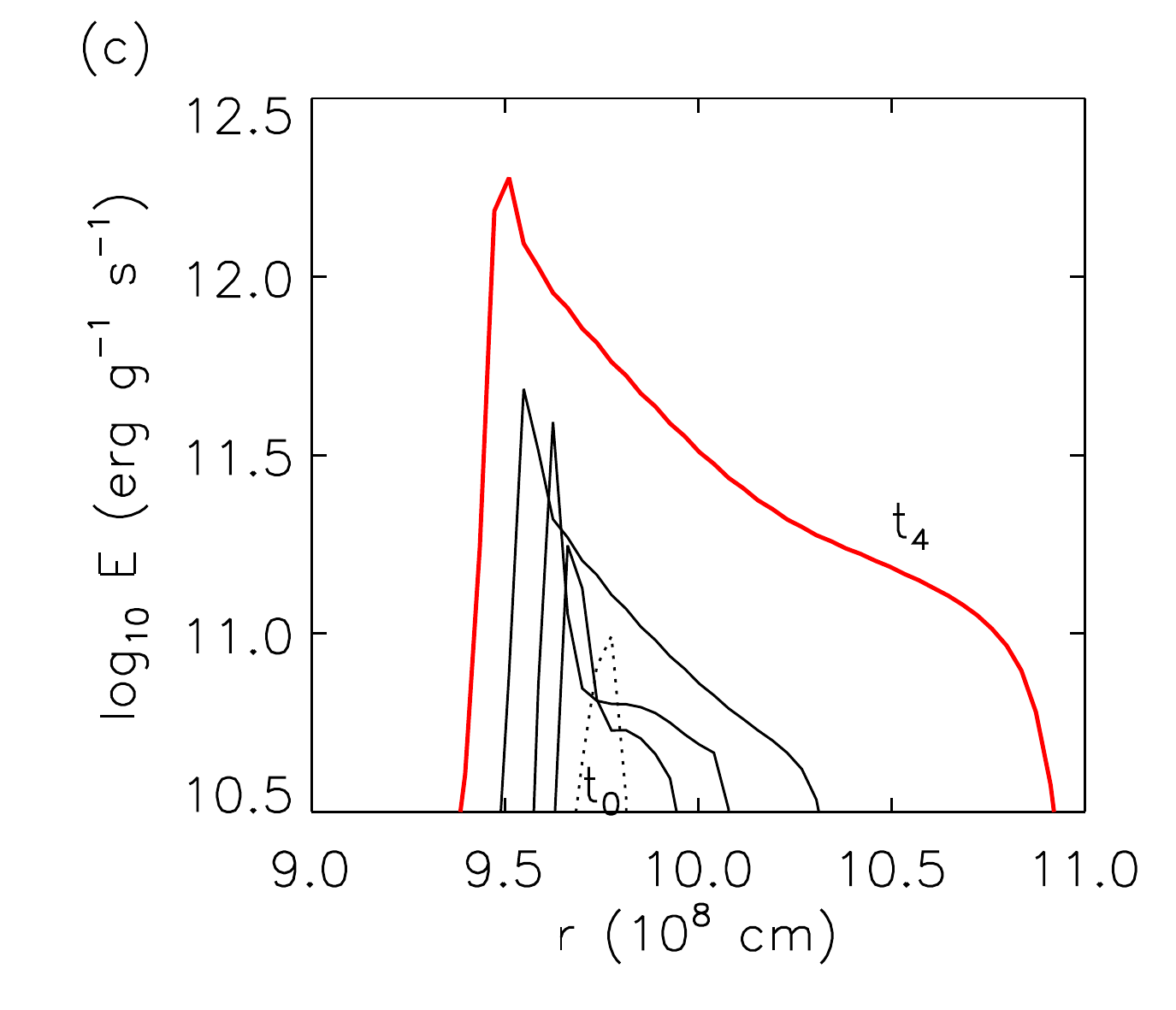}
\caption{Radial distribution of (a) temperature, (b) hydrogen mass
  fraction, and (c) energy production rate at the H-He interface for
  the 3D model hifexp.3d.f at five different times: $t_0 =0\,$s, $t_1
  = 3230\,$s, $t_2 = 6500\,$s, $t_3 = 10\,400\,$s, and $t_4 =
  14\,340\,$s. }
\label{3dsplit}
\end{figure*}

Nuclear burning at the H-He interface steadily increases during both
simulations, and eventually leads to the formation of a distinct
second temperature peak (Fig.\,\ref{fig.temph1entr},
Fig.\,\ref{3dsplit}). This behavior is well known from 1D stellar
evolutionary calculations of the core helium flash in extremely
metal-poor stars, where the helium-burning convection shell splits
after the injection of hydrogen. However, the splitting process seen
in 1D stellar calculations differs from that found in our
multidimensional simulations.  In 1D simulations the splitting occurs
due to the appearance of a temperature peak in the convective
helium-burning shell at the location where hydrogen burning by the
$^{12}\mbox{C} (\mbox{p}, \gamma)^{13}\mbox{N}$ reaction is proceeding
the fastest.

\begin{figure*} 
\includegraphics[width=0.49\hsize]{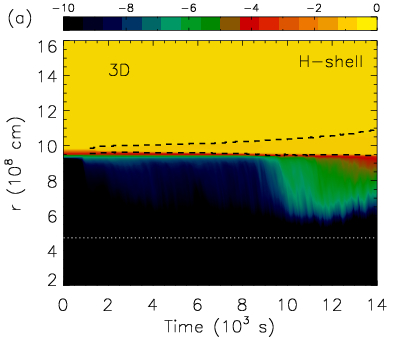}
\includegraphics[width=0.49\hsize]{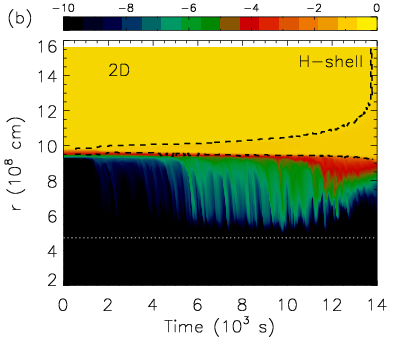}
\caption{(a) Temporal evolution of the angle-averaged radial
  distribution of (the logarithm of) the hydrogen mass fraction for
  the 3D model hifexp.3d.f, and (b) for the 2D model hifexp.2d.f,
  respectively. The dashed lines delineate the boundaries of the newly
  forming hydrogen-burning convection shell, and the dotted line marks
  the bottom of the helium-burning convection zone.}
\label{fig.h1nentr}
\end{figure*}

In our hydrodynamic models the appearance of a second temperature peak
is preceded by a slow migration of the H-He boundary into the
helium-burning convective shell, which forces it to retreat slightly
(Fig.\,\ref{3dsplit}). Eventually, when the nuclear energy release and
the temperature at the H-He boundary become sufficiently high, we
observe the appearance of a new convection shell in the hydrogen-rich
layers above the original one. Hence, what actually happens is not a
splitting, but a ``retreat'' of the initial helium-burning convection
shell and the birth of a new one on top of it powered by burning of
hydrogen.

Our hydrodynamic simulations exhibit another major distinct property:
the absence of an impermeable radiative layer between the two
convective shells. Mixing of nuclear species between these layers is
not prohibited, but occurs during the whole event
(Fig.\,\ref{fig.h1nentr}) with a decreasing efficiency towards the end
of simulation.

The evolution of the hydrodynamic models is likely representative of
early phases during the dual core and He-shell flashes, and indicates
that the entropy barrier does not inhibit hydrogen
dredge-down. However, this conclusion should be moderated by the fact
that the entropy at the H-He boundary can increase by the rise of the
nuclear energy production.  Eventually, this entropy barrier may
become so high that mixing ceases completely. What we have shown here
is that, given our initial conditions, nothing seems to prevent
H-injection, at least in the early stages. A similar behavior was also
found by \citet{Herwig2011} in their 3D simulations of the hydrogen
injection into the helium shell flash convection zone which helped
them to explain major features observed in spectra of the Sakurai's
object \citep{Duerbeck2000} (post-AGB object experiencing its last
helium flash). Finally note that proton injection may have interesting
consequences concerning the production of s-process elements as some
$^{13}$C can be produced which is a source of neutrons
\citep{Campbell2010}.

We note, that during its final evolution the 2D model hifexp.2d.f
eventually experiences another unrealistic nuclear runaway when He
ignites in the secondary convection shell (due to high temperature
$\gtrsim 10^8$~K at the H-He boundary) causing it to rapidly expand to
the outer boundary of the computational domain, when we stopped the
calculation (Fig.\,\ref{fig.h1nentr}~b).  Concerning this unexpected
behavior we have to emphasize that these hydrodynamical simulations
are strongly dependent on the initial conditions. In particular, we
note that in the initial model MOD the layers at the H-He boundary are
highly degenerate (degeneracy parameter $\psi \approx 10$).
\footnote{Matter at the H-He boundary after proton injection (and some
  earlier expansion) in a 1D population III model of 0.85 \Msun
  computed by Simon Campbell (private communication)
  \citep{Mocak2009phd} has a degeneracy parameter $\psi \approx -2$.}
Such a large degeneracy favors a nuclear runaway.  Additionally, our
multidimensional simulations are likely to underestimate the star's
expansion and subsequent cooling because of the reflective boundary
conditions imposed in the radial direction. These numerical restrictions
limit the cooling and favor the aforementioned helium runaway.

\subsection{Simulation with partial CNO network}
\label{sect.simpartcno}

In the 2D simulation hifexp.2d.p we did neither consider the
$\beta$-decay reactions $^{13}\mbox{N}(\beta^-)^{13}\mbox{C}$ and
$^{15}\mbox{O}(\beta^- )^{15}\mbox{N}$, nor the
$^{12}\mbox{C}(\mbox{p}, \gamma)^{13}\mbox{N}$ reaction, which gives
rise to a reduced energy production rate by roughly a factor of three
at the H-He interface.  It allowed us to demonstrates, that entropy
barriers at convection zone boundaries that are insufficiently
sustained by nuclear burning are penetrable and unable to prevent the
growth of the convection zone due to turbulent entrainment on dynamic
timescales. By neglecting proton capture on $^{12}\mbox{C}$ in the
simulation, we obtain an incorrect hydrogen profile and a reduced
nuclear energy production rate. That explains why we do not observe
here the formation of the secondary hydrogen-burning shell convection
as in our hydrodynamic models with the full nuclear CNO reaction
network (see previous subsection).

In this model, helium-burning convection starts at $t \sim 1000\,$s
and extends throughout the whole initially convectively unstable shell
as determined by the Schwarzschild criterion. The convection shell is
located between the radius of the temperature maximum and the base of
the hydrogen-rich layers. It is characterized by vortices, which are
typical structures of 2D simulations due to the imposed axial
symmetry. Their width is equal to the radial extent of the convection
shell, \ie $r\sim 5 \times 10^8\,$cm, which is also the width of the
convection zone in the unmodified initial model (M).

\begin{figure} 
\includegraphics[width=0.99\hsize]{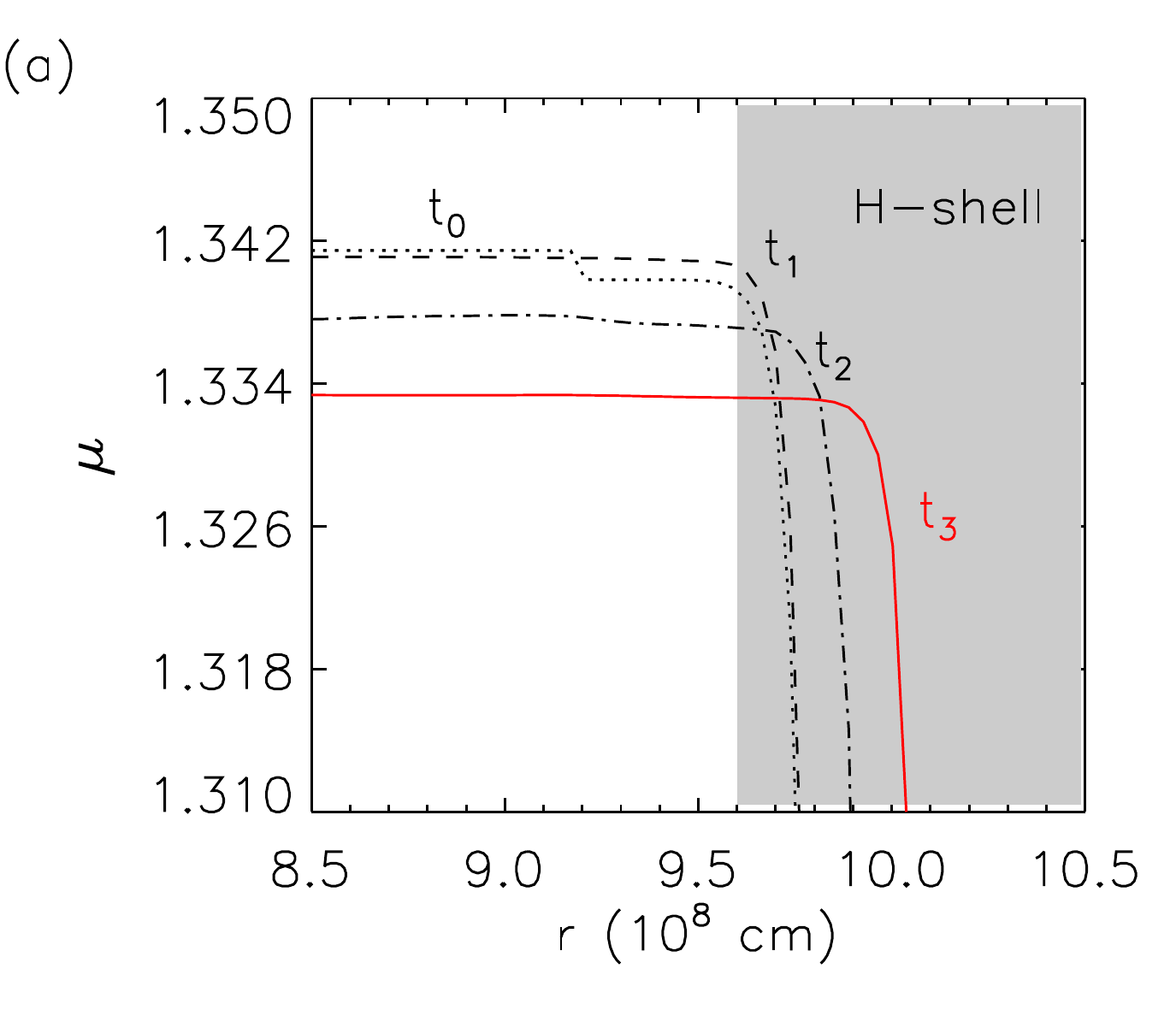}
\includegraphics[width=0.99\hsize]{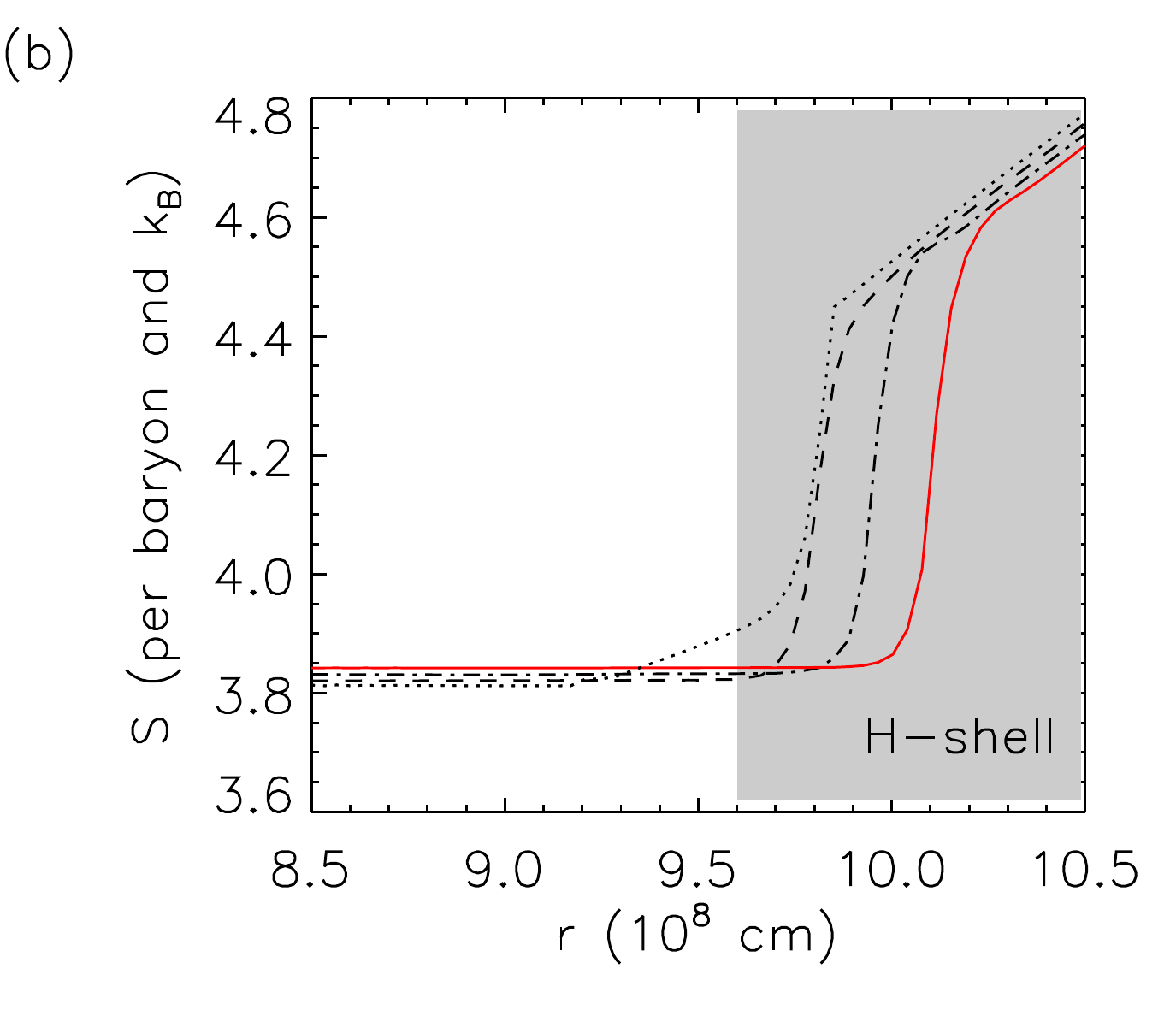}
\caption{Mean molecular weight (a) and entropy (b) at the outer edge
  of the convection zone of the 2D model simulated with the partial
  CNO network (hifexp.2d.p) at four different times: $t_0 = 0\,$s
  (dotted), $t_1 = 46\,600\,$s (dashed), $t_2 = 94\,200\,$s
  (dash-dotted), and $t_3 = 142\,000\,$s (solid red).}
\label{fig.hydent}
\end{figure}

The shell convection reaches the hydrogen-rich layers shortly after
the appearance of the first convective plumes. It initiates the
dredge-down of protons deep into the helium core, and causes the outer
boundary of the convection shell to cross the entropy barrier at the
H-He interface (Fig.\,\ref{fig.hydent}). The value of the entropy at
this interface constantly decreases due to turbulent entrainment,
which effectively shifts the entropy barrier to larger radii.

The entrainment rate of the outer convection boundary is initially
relatively high, around $70 \mes$, but after roughly 30\,000\,s it
drops and eventually relaxes to an almost constant value of $\sim 4.5
\mes$.  The entropy barrier has a stabilizing effect because it
decreases the entrainment velocity. However, it does not prevent the
advance of the convection zone into the H-rich layers and leads to an
erosion of the entropy gradient. After 170\,000\,s the shell
convection has moved $\sim 8\times 10^7\,$cm into the H-rich buffer,
meaning that almost 12\% of the hydrogen-rich shell included into the
simulation (limited by the outer radial grid boundary at $1.6\times
10^9\,$cm) is already entrained within only 47 hours, or $\sim 300$
convective turnover timescales.

In this simulation, protons are able to reach layers with temperatures
as high as $\sim 1.8\times 10^8\,$K. However, such a deep mixing of
hydrogen into the helium burning layers is not realistic because
protons should have been captured by $^{12}\mbox{C}$ before.

\section{Stellar evolutionary calculations}
\label{sect:stec}

To mimic the effects of the turbulent entrainment in 1D stellar
evolutionary calculations, we experiment and implement a simple
entrainment law
\begin{equation}
  \dot{M}_E = \frac{\partial M}{\partial r} u_E
\end{equation}
where $M$ is the mass of the shell at radius $r$ and $u_E$ is the
entrainment rate \citep{MeakinArnett2007}. We simulated a $1\,\Msun$,
$Z=Z_\odot$ star up to the core helium flash. At first we use an
entrainment rate of 6.7$\times 10^2 \cms$ as found by \cite{Mocak2009}
in their 3D hydrodynamic simulations of the core helium flash.  The
entrainment is switched on only shortly before the flash reaches its
maximum intensity (maximum nuclear energy generated in the helium
core). We find that the He-convection zone boundary is unable to cross
the entropy barrier and penetrate into the hydrogen shell. This result
also holds when increasing the entrainment rate by one or even two
orders of magnitude.

This is partially due to the inherent time dependence of the mixing
prescription. Indeed, for a given timestep $\Delta t$, mixing will
extend over a mass range given by $\Delta M_{e} = \dot{M}_E \Delta
t$. Practically, we simply shift the convective boundary to the new
mass location $M_{conv}^{edge}+\Delta M_{e}$. As the helium convection
shell approaches the H-rich layer, the timestep which is constrained
among other things by the rate of change of the nuclear luminosity,
decreases thus preventing the injection of protons and a subsequent
H-flash. Eventually, a (small) timestep is found that allows the
simulation to go on and after the some time, the helium-shell
convection zone naturally quenches and H-injection is avoided.

On the other hand, if we experiment with overshooting by using the
exponential decay of the convection velocity in the radiative layers,
mixing is treated as a diffusive process with the diffusion
coefficient given by
\begin{equation}
D_{over} = D_{conv}\exp(-|r - r_{edge}|/(2f H_P)
\end{equation}
where $H_P$ is the pressure scale height at the edge of the convective
zone and $f$ a free parameter. Compared to the previous prescription,
this treatment shows two main differences : mixing is treated as a
diffusive process so the penetration of the He-shell convection is
``smoother'' and the new convective boundary at $t+\Delta t$ does not explicitly
depend on its location at time $t$ and on the timestep. Our 1D experiments 
indicate that at solar
metallicity, a substantial overshooting is required to trigger an
injection with $f\sim 1$. This injection then leads to the splitting
of the He-shell convection and after some structural readjustment to
the pollution of the envelope. We report C and N enrichments by a
factor of 2 or 3. Interestingly, we find that subsequently to this
mixing event, the \chem{12}C/\chem{13}C ratio has dropped to below
$\sim 10$. Moreover, these He-core burning stars are carbon-rich
(C/O$\,\sim 1.7$) and show large \chem{7}Li abundances
($\varepsilon(Li) \approx 3.5$). Some of these features are found in
J-stars but contrarily to observations, our models show a too large
\chem{16}O/\chem{17}O ratio (by a factor of two at least, and the C/N
ratio is too low in our models).

\section{Summary}
\label{sect:sum}

We performed hydrodynamic simulations of shell convection in the core
of a low mass star at the peak luminosity of the helium flash. In
these computations, the hydrogen-rich layers were artificially shifted
down into the helium core. They allowed us to demonstrate a
penetration of the helium-burning shell convection into the hydrogen
layers due to turbulent entrainment despite the presence of an entropy
barrier between them. Resulting consequences are the following.

Hydrodynamic models computed with the full CNO reaction network leads
to the formation of a secondary shell convection zone in the
hydrogen-rich layers. It is powered by hydrogen burning and is located
above the already-present helium-burning shell convection. These
simulations show (contrary to 1D stellar calculations of similar
evolutionary phases) that the mixing of protons down to the
helium-burning convection shell is not completely inhibited, although
it eventually slows down due to an increase of the nuclear energy
production and the growth of the entropy barrier at the interface
between the helium-burning and hydrogen-burning convection shell.

In our hydrodynamic model computed with a reduced nuclear reaction
network the energy production rate in the hydrogen-rich layers is
smaller and we do not observe an appearance of the hydrogen-burning
convection shell.  This simulation shows, that an entropy barrier 
  not sustained sufficiently by nuclear burning does not act as an
obstacle for the growing helium-burning convection shell due to
turbulent entrainment. The entrainment rate of the outer rim of the
helium-burning convection zone reaches a roughly constant
non-vanishing value.

Given our initial model of low-mass and metal-rich star, the
aforementioned situations were difficult to reconstruct by
one-dimensional stellar evolution code.

\begin{acknowledgements}
The simulations were performed at the Rechenzentrum Garching on the
IBM Power6 system. The authors want to thank Frank Timmes for some of
his publicly available FORTRAN subroutines which we used in the
HERAKLES code. Miroslav Moc\'ak acknowledges financial support from
the Communaut\'e fran\c caise de Belgique - Actions de Recherche
Concert\'ees. LS is FNRS research associate. We also thank John
Lattanzio for summarizing the observations about J-stars.
\end{acknowledgements}

\bibliography{referenc}

\end{document}